\documentclass[sigconf, natbib=false]{acmart}
\usepackage[style=ACM-Reference-Format, maxbibnames=3, backend=bibtex, sorting=none]{biblatex}

\usepackage{pifont}
\usepackage{amsmath}
\usepackage{balance}
\usepackage{threeparttable}
\usepackage{xcolor}
\usepackage{soul}
\usepackage{subcaption}
\usepackage[linesnumbered, ruled, vlined]{algorithm2e}
\let\oldnl\nl
\newcommand{\nonl}{\renewcommand{\nl}{\let\nl\oldnl}}
\usepackage{comment}

\usepackage{array}
\usepackage{algcompatible}

\usepackage{multirow}
\usepackage{graphicx}
\usepackage{color, soul}
\usepackage{color, colortbl}

\usepackage{makecell}

\usepackage{breqn}

\usepackage{multirow}
\usepackage{breqn}
\AtBeginDocument{%
  \providecommand\BibTeX{{%
    \normalfont B\kern-0.5em{\scshape i\kern-0.25em b}\kern-0.8em\TeX}}}

\setcopyright{none}
\renewcommand\footnotetextcopyrightpermission[1]{}

\usepackage{tikz,tkz-euclide,pgfplots}
\tikzstyle{roundnode} = [circle, fill=black!255, scale=.8]

\tikzstyle{squarenode} = [square, fill=black!255, scale=1.0]

\begin{document}
\fancyhead{}

%
\title{Heterogeneous Integration of In-Memory Analog Computing Architectures with Tensor Processing Units}

\author{Mohammed E. Elbtity}
\email{elbtity@ieee.org}
\orcid{0000-0002-3282-0076}
\affiliation{%
  \institution{University of South Carolina}
  \city{Columbia}
  \state{SC}
  \country{US}
  \postcode{29205}
}

\author{Brendan Reidy}
\email{bcreidy@email.sc.edu}
\orcid{0009-0004-4320-6890}
\affiliation{%
  \institution{University of South Carolina}
  \city{Columbia}
  \state{SC}
  \country{US}
  \postcode{29205}
}

\author{Md Hasibul Amin}
\email{ma77@email.sc.edu}
\orcid{0000-0002-9919-6626}
\affiliation{%
  \institution{University of South Carolina}
  \city{Columbia}
  \state{SC}
  \country{US}
  \postcode{29205}
}

\author{Ramtin Zand}
\email{ramtin@cse.sc.edu}
\orcid{0000-0002-1786-1152}
\affiliation{%
  \institution{University of South Carolina}
  \city{Columbia}
  \state{SC}
  \country{US}
  \postcode{29205}
}

\renewcommand{\shortauthors}{Mohammed E. Elbtity, Brendan Reidy, Md Hasibul Amin, \& Ramtin Zand}


\begin{abstract}

Tensor processing units (TPUs), specialized hardware accelerators for machine learning tasks, have shown significant performance improvements when executing convolutional layers in convolutional neural networks (CNNs). However, they struggle to maintain the same efficiency in fully connected (FC) layers, leading to suboptimal hardware utilization. In-memory analog computing (IMAC) architectures, on the other hand, have demonstrated notable speedup in executing FC layers. This paper introduces a novel, heterogeneous, mixed-signal, and mixed-precision architecture that integrates an IMAC unit with an edge TPU to enhance mobile CNN performance. To leverage the strengths of TPUs for convolutional layers and IMAC circuits for dense layers, we propose a unified learning algorithm that incorporates mixed-precision training techniques to mitigate potential accuracy drops when deploying models on the TPU-IMAC architecture. The simulations demonstrate that the TPU-IMAC configuration achieves up to $2.59\times$ performance improvements, and $88\%$ memory reductions compared to conventional TPU architectures for various CNN models while maintaining comparable accuracy. 
The TPU-IMAC architecture shows potential for various applications where energy efficiency and high performance are essential, such as edge computing and real-time processing in mobile devices. The unified training algorithm and the integration of IMAC and TPU architectures contribute to the potential impact of this research on the broader machine learning landscape.

\end{abstract}

\maketitle
\pagestyle{plain}

\section{\textbf{Introduction}}

Deep learning models have been widely adopted in various real-life applications, including language translation, computer vision, healthcare, and self-driving cars \cite{Hemed2018DistributedDL, Zhou2020APH, Shrivastava2019DeepLF, Miotto2018DeepLF}. However, this has resulted in a significant increase in the computational demands of machine learning (ML) workloads, which conventional von Neumann architectures struggle to keep up with. To overcome this challenge, alternative architectures such as in-memory computing (IMC) have emerged, which perform computations directly where the data exists, thus reducing the high energy costs of data transfers between memory and processor in data-intensive applications like ML \cite{Sebastian2020MemoryDA}. 
%
%
Conventional IMC architectures typically employ emerging technologies such as resistive random access memory (RRAM) \cite{Shaarawy2018DesignAA,Yin2020HighThroughputIC} and magnetoresistive random-access memory (MRAM) \cite{inmemory-fan} to accelerate matrix-vector multiplication (MVM) operations in the ML workloads through massive parallelism and analog computation. However, other functional blocks such as activation functions still rely on digital computation, resulting in energy overheads due to signal conversion units \cite{PUMA,dpengine}. In-memory analog computing (IMAC) architectures, on the other hand, are a class of IMC architectures, which realize both MVM operations and non-linear vector operation in the analog domain, and thus obviate the need for signal conversion units between deep neural networks (DNNs) layers \cite{Amin2022MRAMbasedAS}. Previous works have shown that IMAC architectures can achieve orders of magnitude reduction in latency and energy consumption in implementing dense fully connected (FC) layers in DNNs \cite{Elbtity2021AnIA}. However, adapting IMAC architectures to implement convolutional layers in convolutional neural networks (CNNs) requires unrolling and reshaping the layers to MVM \cite{xbar-unroll}, resulting in large crossbar arrays that may be susceptible to reliability issues caused by noise and interconnect parasitic \cite{Amin2022XbarPartitioningAP,iscas-imac}.

On the other hand, one of the most promising digital hardware accelerators introduced in recent years to accelerate ML workloads is tensor processing units (TPUs), which performs parallel computing with a deeply-pipelined network of processing elements (PEs), namely, systolic arrays \cite{jouppi2017datacenter}. Systolic arrays in TPUs reduce energy consumption and increase performance by reusing the values fetched from memory and registers and reducing irregular intermediate memory accesses \cite{TPU-micro2018}. 
TPUs have demonstrated impressive results in executing the general matrix multiplication operation, which is a critical component of CNNs, specifically in convolutional layers. However, TPUs struggle to maintain the same level of performance when executing FC layers due to the vast number of weights that typically make up FC layers. This limits weight reuse and necessitates multiple iterations to execute, resulting in inefficient hardware utilization and high energy consumption \cite{Ravikumar2022EffectON}. 
It shall be noted that in practice, systolic arrays have a symmetrical size to optimize the performance of Convolutional layer execution. However, if designed with asymmetric dimensions, they can accelerate FC operations at the cost of convolutional layer execution performance. 
Recent research \cite{Ravikumar2022EffectON} has shown that TPUs outperform GPUs and CPUs in executing convolutional layers due to their spatial reuse characteristics and ability to drop unused features in the neural network. However, TPUs are less preferred for running FC layers and perform worse than GPUs due to lower weight reuse. This leads to high memory traffic and energy consumption, particularly as the model size increases. The percentage of FC layers in a given neural network architecture impacts the overall performance of executing the entire model on the TPU. Our in-house experiments using Scale-Sim \cite{samajdar2018scale} also confirm poor performance and inefficient hardware utilization of TPUs when executing FC layers compared to convolutional layers.

Here, we present a novel hybrid TPU-IMAC architecture that combines the strengths of TPU and IMAC to efficiently execute both Convolutional and FC layers. This architecture results in performance improvements, as well as reduced memory bandwidth requirements for various-sized CNN models. The paper is structured as follows: Section 2 describes the IMAC architecture used in this study, while Section 3 presents a detailed description of the proposed TPU-IMAC architecture. In Section 4, we explain our architecture-aware learning method for training mixed-precision CNN models. Section 5 discusses the accuracy, memory utilization, and performance results. Finally, Section 6 concludes the paper.

\section{IMAC Architecture}



\begin{figure}[!t]
\centering
\includegraphics[width=2.5in]{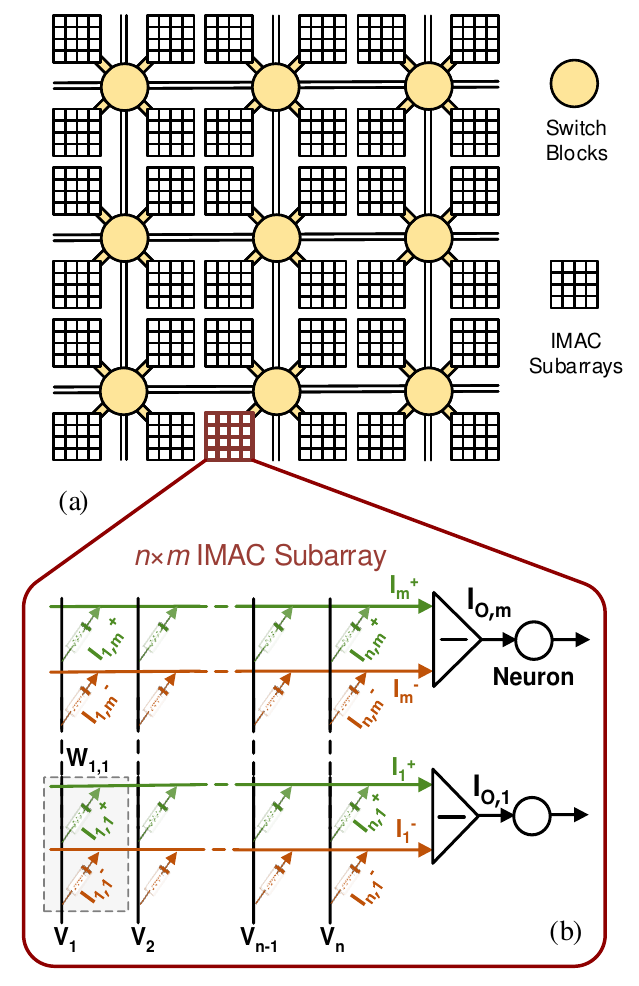}
\vspace{-3mm}
\caption{(a) IMAC architecture, (b) An $n \times m$ subarray including memristive crossbars and analog neurons.}
\label{fig:arch}
\end{figure}

The diagram presented in Figure 1 illustrates the structures of IMAC architectures. These architectures consist of a set of closely interconnected subarrays, linked by programmable switch blocks. Each IMAC subarray is made up of memristive crossbars, differential amplifiers, and neuron circuits, as depicted in Figure \ref{fig:arch} (b). 
For the sake of simplicity, we exclusively illustrate the read path of the arrays, as the focus of this study is on the inference phase of the neural network. The synaptic connections of the DNN are created by the memristive crossbars, which have a number of columns and rows that can be defined based on the number of inputs and output nodes in a single FC layer, respectively.
The memristor crossbars execute the MVM operation in the analog domain using physical mechanisms like Ohm's law and Kirchhoff's law in electrical circuits. Specifically, the multiplication operation is performed according to Ohm's law ($I = GV$), while the accumulation operation is based on the conservation of charge, as explained by Kirchhoff's law.


During the configuration phase, adjusting
the relative conductance of two memristive devices connected to a differential amplifier enables the realization of zero, positive and negative weights. Figure 1 (b) illustrates that the differential amplifiers are linked to two adjacent rows in the crossbar that are colored green and red, representing positive and negative rows of conductances, respectively. The differential pair of memristive devices with conductance values of $G_{i,j}^+$ and $G_{i,j}^-$ is used to realize each weight value $W_{i,j}$, where $W_{i,j} \propto G_{i,j}^+-G_{i,j}^-$. Thus, a pair of memristive devices having $G_{i,j}^+=1/R_{high}$ and $G_{i,j}^-=1/R_{low}$ is used to implement negative weight and vice versa. The zero weight is realized if $G_{i,j}^+=G_{i,j}^-$.



During the inference phase, the write word lines (WWLs) are disabled, and the read word lines (RWLs) are enabled. This process generates two types of currents, $I^+$ and $I^-$, as shown in Figure \ref{fig:arch} (b), with the current amplitude depending on the input signals and the resistances of the memristive synapses. Each row of memristive synapses shares a differential amplifier that produces an output voltage proportional to the difference between the currents of the two word lines for that row, i.e., $\sum_{i}(I_{i,n}^+-I_{i,n}^-)$, where $i$ is the total number of nodes in the input layer, and $n$ is the row number. Finally, the output of the differential amplifiers is fed to the analog neurons to compute the activation functions. This architecture performs both MVM operations and neuron activation functions in each subarray for a given layer and then passes the result to the next IMAC subarray to compute the next layer. In this work, we use an analog \textit{sigmoid} neuron, which is composed of two resistive devices and a CMOS-based inverter \cite{Amin2022MRAMbasedAS}. The resistive devices in the neuron's circuit form a voltage divider that reduces the slope of the inverter's linear operating region, resulting in a smooth high-to-low voltage transition that creates a sigmoid function.





\begin{figure*}[]
\centering
\includegraphics[width=4.6in]{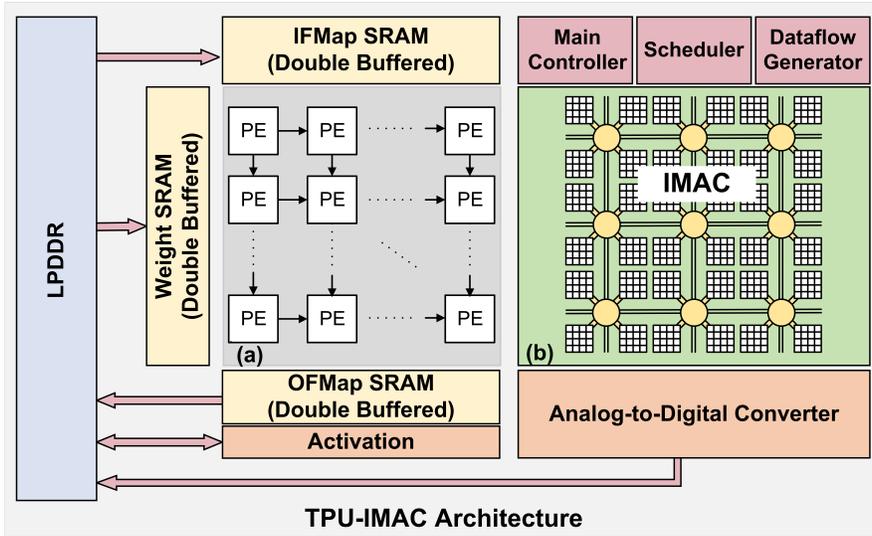}
\vspace{-2mm}
\caption{Heterogeneous TPU-IMAC architecture. (a) Output stationary systolic array including a deeply-pipelined network of processing elements (PEs), and (b) a network of tightly-coupled memristive subarrays interconnected with switch boxes.}
\label{fig:tpu-imac}
\vspace{-3mm}
\end{figure*}

\section{Proposed TPU-IMAC Architecture.}



Generally, the TPU architecture encompasses a multitude of PEs that include multiply-and-accumulate (MAC) units responsible for executing matrix-matrix, vector-vector, and matrix-vector multiplications\cite{Elbtity2022APTPUAC}. The TPU leverages the systolic array to enhance performance by reusing values retrieved from memory and registers \cite{TPU-micro2018}, consequently minimizing reads and writes to buffers. Input data is concurrently fed into the array, usually propagating in a diagonal wavefront pattern \cite{jouppi2017datacenter, shen2019high}. The fundamental architecture of a MAC unit influences data flow within the systolic array, and  varying data flow architectures impact power consumption, percentage of hardware utilization, and overall performance.

Data flow in the systolic array for neural network processing is deliberately arranged to extract data and generate output results in a deterministic sequence that optimizes MACs, the primary operations in deep learning algorithms. A variety of  data flow algorithms have been proposed and can be broadly categorized as input stationary data flow (IS) \cite{lym2020flexsa}, weight stationary (WS) \cite{jouppi2017datacenter}, output stationary (OS) \cite{guo2019systolic}, row stationary (RS) \cite{chen2016eyeriss}, and no-local reuse (NLR) \cite{qin2020sigma, lim2000serial}.
The term "stationary" denotes the data that remains within the PEs and does not travel through the registers while carrying out MAC operations \cite{samajdar2018scale}. In the WS approach, each weight is pre-loaded onto a MAC unit. During each cycle, input elements to be multiplied by the affixed weights are broadcasted across the MAC units in the array, yielding partial sums at every clock cycle. This procedure is distributed vertically over columns to generate results from the bottom of the systolic array \cite{jouppi2017datacenter}. A nearly identical process occurs for IS data flow, where inputs remain constant in the matrix while weights are disseminated horizontally. For OS data flow, outputs are attached to MAC units as inputs and weights are circulated among the units. Figure \ref{fig:tpu-imac}(a) illustrates the architecture of an OS systolic array, where the weights are introduced from the left side of the array and the input feature map (IFMap) is streamed in from the top. Each PE is accountable for generating an output feature map (OFMap).

We propose a hybrid-precision, mixed-signal, and heterogeneous technology system that integrates resistive RRAM-based IMAC with a conventional CMOS-based TPU architecture. The proposed TPU-IMAC architecture, illustrated in Figure \ref{fig:tpu-imac}, retains the conventional TPU functionality while enhancing overall performance by incorporating IMAC circuits directly connected to the PEs within the TPU's systolic arrays. We adhere to the OS architecture in the systolic array, capitalizing on its ability to fix OFMaps data in corresponding PEs. This enables direct connection of the most significant bit ($sign$ bit) of each OFMap to the IMAC inputs, facilitating the immediate transfer of Convolutional layer results from the TPU to the IMAC for executing the subsequent FC layer, depending on the neural network topology.
Data quantization occurs without the need for specialized hardware or software functions by connecting the $sign$ bit through an inverter, converting positive OFMaps ($\geq0$) to a high logic bit '1' and negative OFMaps (high $sign$ bit) to a low logic bit '0'. This single-bit precision data is connected to the IMAC inputs via a tri-state buffer controlled by the \textit{Main Controller} during FC layer execution on the IMAC.

In the proposed architecture, we employ low-power double data rate (LPDDR) memory, which is suitable for edge devices due to its lower operating voltage and power-saving modes. LPDDR is responsible for storing and retrieving neural network IFMap, weight, and OFMap data according to the \textit{dataflow generator} and \textit{Main Controller}. Typically, data is pre-loaded into LPDDR, and when the TPU-IMAC begins workload execution, the \textit{dataflow generator} generates read address traces for retrieving IFMaps and weights from LPDDR, sending them to IFMap SRAM and weight SRAM, respectively, based on the OS dataflow algorithm. The \textit{Main Controller} facilitates this data transfer, following the \textit{scheduler} request. 

After executing the first Convolutional layer, OFMap data is forwarded from the PEs to the OFMap SRAM and then transferred to LPDDR according to the dataflow and write traces (addresses) from the \textit{dataflow generator}.
The \textit{scheduler} is responsible for scheduling each layer of the CNN workload, while the \textit{dataflow generator} and \textit{Main Controller} manage the overall flow of CNN workload execution. Depending on the CNN workload, the \textit{scheduler} may need to execute one or more FC layers. In this case, the \textit{scheduler} informs the \textit{Main Controller}, which enables data movement between the $sign$ signals of each OFMap (stored in each PE of the systolic array) and the IMAC inputs by activating the in-between tri-state buffers. 
The input data of the FC layers are in low or high logic form, while IMAC weights utilize ternary logic (with values 1, 0, and -1). 

We train the CNN workload on the proposed architecture to mitigate potential accuracy loss resulting from low-precision analog computation. Once the data moves from the systolic array to the IMAC, IMAC executes the required FC layers based on the \textit{scheduler}'s request, with each FC layer executed in a single clock cycle, improving performance. It is important to note that the FC weights are pre-loaded onto the memristive devices within the IMAC circuits in the configuration phase as described in section 2. Upon completing the FC layer execution on the IMAC, the results are converted to digital using the attached analog-to-digital converter (ADC) within the IMAC architecture, and then written back to LPDDR for user access.

It is noteworthy that our architecture does not require a digital-to-analog converter (DAC) since our IMAC accepts binarized inputs that are coming directly from the $sign$-bit of each PE in the systolic array, resulting in reduced power consumption.
If an activation or normalization layer is required, a specialized hardware unit is implemented outside the TPU's systolic array to perform these operations accordingly.
Figure \ref{fig:tpu-imac} also depicts the dataflow between the architecture components using arrows for simplification.

In summary, the \textit{scheduler} controls the execution of each layer and is programmed according to the CNN topology. The \textit{dataflow generator} generates traces (addresses) for LPDDR to read data and send it to the IFMap memory and weight memory, or to write results from the OFMap memory or the ADC to LPDDR based on the OS dataflow algorithm. The \textit{Main Controller} manages the enable signals of each component and the tri-state buffers between the TPU's systolic arrays and the IMAC circuits. It is worth noting that each PE within the TPU's systolic array contains a full-precision 32-bit floating-point (FP) MAC unit, while the IMAC utilizes ternary weights and binary inputs as already explained. This unique combination of precision and mixed-signal technology within the proposed TPU-IMAC architecture offers an innovative approach to enhancing CNN inference performance. 

\section{Architecture-Aware CNN Model Development}

We developed a hardware-aware learning algorithm to fully exploit the advantages of the TPU-IMAC architecture while maintaining accuracy. Our mixed-precision and mixed-signal TPU-IMAC architecture has computational constraints and unique features, which our algorithm takes into account. The learning process consists of two steps. In the first step, we insert a $tanh$ activation function before the FC section of the network to ensure that activations stay within the range of \{\textit{-1, 1}\}. After that, we train the CNN using the backpropagation algorithm. In the second step, we freeze the trained convolutional layers and retrain the FC layers using ternary weights. Here, we replace the $tanh$ activation function with the $sign$ function to produce input values of \textit{-1} and \textit{1} for the dense layers. This is important because by restricting the inputs of the dense layers to \textit{-1} and \textit{1}, we only need to transfer the sign bit of the last convolution layer's OFMaps to IMAC, without requiring any DAC units. Additionally, we modify the FC layers based on the characteristics of IMAC, employing ternary synapses and $sigmoid$ activation functions that can be realized using RRAM-based synapses and neurons, as described in Section 2.

\begin{table}[]
\caption{The weights and activations in different stages of the proposed TPU-IMAC-aware learning algorithm.}
\vspace{-3mm}
\begin{tabular}{cclccc}
\hline 
Step               & layers                        & \multicolumn{1}{c}{Component} & \multicolumn{1}{c}{\begin{tabular}[c]{@{}c@{}}Forward\\ Pass\end{tabular}} & \multicolumn{2}{c}{\begin{tabular}[c]{@{}c@{}}Backward\\ Pass\end{tabular}} \\ \hline \hline
\multirow{2}{*}{1} & \multirow{2}{*}{All} & Weights                       &       $\textbf{w}_i \in R$                                                                     &  $\textbf{w}_i \in R$                                                      \\
                   &                               & Neuron                        &           ReLU                                                                 &  ReLU                                                      \\ \hline
\multirow{4}{*}{2} & \multirow{2}{*}{Conv}  & Weights                       & \multicolumn{3}{c}{\textit{Frozen}}                                                                                                                               \\
                   &                               & Act.                          & \multicolumn{3}{c}{ReLU}                                                                                                                                 \\ \cline{2-6} 
                   & \multirow{2}{*}{FC}        & Weights                       &        $\textbf{W}_i \in \{-1, 0, +1\}$                                                                    & $\textbf{w}_i \in R$                                                        \\
                   &                               & Neuron                        &         \textit{sigmoid}                                                                   & \textit{sigmoid}                                                            \\ \hline
\end{tabular}
\label{tab:learning}
\end{table}

In Table \ref{tab:learning}, we present the activation functions and precision of weights in convolutional and dense layers for each training step mentioned earlier. In forward pass of the second step of training, the FC layers are trained using ternary weights, while in the backward pass, FP weights are used. After retraining the model with ternary weights, only the ternary weights are kept. It is worth noting that most existing CNN models use ReLUs to achieve a non-saturating nonlinearity because of their implementation simplicity and performance benefits compared to digital implementations of $tanh$ and $sigmoid$ activation functions. However, in our proposed IMAC architecture, our analog neurons realize high-performance sigmoidal activation functions, which provide accuracy benefits with minimal performance overheads. Although ReLU is still used in the convolutional layers implemented on the TPU, in IMAC architecture, we use analog sigmoidal activation functions. To fully utilize the TPU-IMAC architecture with a $32\times32$ systolic array size, we modify the CNN models to have exactly 1024 elements in the linear vector fed to the dense layer after flattening the last convolutional layer's OFMap. This way, we can directly transfer the OFMap of the last convolutional layer, computed and stored in the systolic array, to IMAC without the need to transfer data to and from the main memory. For VGG9 and ResNet, this is achieved by increasing the number of channels in the final convolutional layer and decreasing the strides on the MaxPooling layer, while for MobileNetV1 and MobileNetV2, this is accomplished by increasing the number of channels in the final convolutional layer.


\begin{table*}[]
\caption{Accuracy, memory utilization and execution time for different CNN models.}
\vspace{-2mm}
\label{tab:absolute}
\centering
\begin{tabular}{l|c|cc|c|ccc|cc}
\hline
\multirow{2}{*}{Model}   & \multirow{2}{*}{Dataset}   & \multicolumn{2}{c|}{Accuracy (\%)}  & \multicolumn{4}{c|}{Memory (MB)} &  \multicolumn{2}{c}{Cycles ($\times10^3$)}  \\ \cline{3-4} \cline{5-8} \cline{9-10}
& & \multirow{2}{*}{TPU}   & \multirow{2}{*}{TPU-IMAC}  & TPU    & \multicolumn{3}{c|}{TPU-IMAC} & \multirow{2}{*}{TPU}   & \multirow{2}{*}{TPU-IMAC}    \\ \cline{5-8}
& &    &   & SRAM    & SRAM & RRAM & Total     \\ \hline
LeNet   & MNIST     & 98.95 & 97.82     & 0.177 & 0.01 & 0.01 & 0.02 &  2.475 & 0.956             \\ \hline
VGG9    & \multirow{4}{*}{CIFAR-10}     & 90.9  & 90.31 &   38.747 & 34.512 & 0.265 & 34.776 & 331 & 297.18 \\
MobileNetV1 && 92.89 & 92.7  & 16.976 & 12.74 & 0.265 & 13.005 & 214.9 & 181.1\\ 

MobileNetV2 && 93.73 & 93.43 & 12.904 & 8.668 & 0.265 & 8.933 & 338.7 & 304.9  \\ 
ResNet-18 && 94.96   & 94.84 & 48.872 & 44.637 & 0.265 & 44.902 & 681.7 & 647.8
\\\hline
MobileNetV1    & \multirow{2}{*}{CIFAR-100}  & 66.21  & 63.07 & 17.344 & 12.74 & 0.288 & 13.028 & 218 & 181.1 \\
MobileNetV2 && 73.06 & 70.14  & 13.272 & 8.668 & 0.288 & 8.956 & 356 & 319.1 \\ 
\hline

\end{tabular}
\end{table*}

\section{Results and Discussion}


\subsection{Accuracy}

\begin{table}[]
\caption{TPU-IMAC accuracy, performance, and memory reduction compared to TPU for various CNN models.}
\vspace{-2mm}
\label{tab:relative}
\centering
\begin{tabular}{lccccc}
\hline
\multirow{2}{*}{Model}     & \multirow{2}{*}{Dataset}   & Accuracy  & Memory    \\
& & Difference   & Reduction  & \multirow{-2}{*}{Speedup}       \\ \hline
LeNet   & MNIST     & -1.13\% & 88.34\%     & 2.59             \\ \hline
VGG9    & \multirow{4}{*}{CIFAR-10}     & -0.59\%  & 10.25\% & 1.11 \\
MobileNetV1 && -0.19\% & 23.39\%  & 1.19 \\ 
MobileNetV2 && -0.3\% & 30.77\% & 1.11 \\ 
ResNet-18 && -0.12\%   & 8.12\% & 1.05
\\\hline
MobileNetV1    & \multirow{2}{*}{CIFAR-100}  & -3.14\%  & 24.89\% & 1.2 \\
MobileNetV2 && -2.92\% & 32.52\%  & 1.12 \\ 
\hline

\end{tabular}
\end{table}







In this study, we conducted experiments on seven different CNN architectures, including LeNet \cite{LeNet} for the MNIST dataset, VGG-9 \cite{VGG-cifar}, MobileNet V1 and V2 \cite{howard2017mobilenets}, and ResNet-18 \cite{resnet} for the CIFAR-10 \cite{CIFAR10} dataset, and MobileNet V1 and V2 for CIFAR-100 dataset, to assess the benefits of using TPU-IMAC architecture over TPU architecture. The models trained for TPU architecture utilized FP32 precision, while TPU-IMAC models are mixed-precision models that incorporated FP32 convolutional layers and ternary dense layers. The accuracy values obtained for both TPU and TPU-IMAC architectures are presented in Table \ref{tab:absolute}. The simulation results indicate a minimal accuracy drop of less than 1\% for the CIFAR-10 dataset for the TPU-IMAC implementation. Specifically, the VGG-9 and ResNet-18 models experienced the maximum and minimum accuracy drop of 0.59\% and 0.12\%, respectively. For the LeNet dataset, the accuracy drop is 1.13\% which can be attributed to its larger ratio of FC-to-Conv layers. Finally, we experience near 3\% accuracy drop for mixed-precision models deployed on TPU-IMAC for CIFAR-100 dataset, which can be due to the complexity of the dataset and larger size of the FC layers compared to those of the CNN models used for CIFAR-10 dataset.


\subsection{Memory Utilization}



Consideration of memory footprint is crucial when deploying ML workloads on edge devices with limited resources \cite{naveen2022memory}. Dense FC layers in CNN models often contribute significantly to memory usage. To address this, utilizing extremely low precision representations, such as ternary weight values represented by only 2 bits, can considerably reduce CNN models' memory usage. Table 2 and Table 3 provide comparisons of memory utilization for single-precision FP models deployed on TPU and mixed-precision models deployed on TPU-IMAC. Simulation results demonstrate that the TPU-IMAC architecture effectively reduces memory usage for the investigated CNN models, thanks to its hybrid memory architecture that integrates conventional SRAM cells with emerging resistive memory technologies like RRAM. Particularly, the TPU-IMAC architecture requires 88.34\%, 18.13\%, and 28.7\% less storage on average compared to TPU for CNN models created for LeNet, CIFAR-10, and CIFAR-100 datasets, respectively.




\subsection{Performance}

In this study, we conducted a performance analysis of the proposed TPU-IMAC architecture using the Scale-Sim simulator. Scale-Sim is a cycle-accurate and architectural-level simulation tool specifically designed for systolic array-based accelerators that execute CNNs \cite{samajdar2018scale}. The tool offers flexible simulation options, including the ability to vary systolic array architecture parameters such as size, dataflow specifications (IS, WS, and OS), as well as DRAM and SRAM sizes, and offsets for the IFMap, filter, and OFMap. By leveraging Scale-Sim, we were able to evaluate the performance of the TPU-IMAC architecture under various configurations and scenarios, providing insights into its potential benefits and limitations for executing CNN workloads on mobile devices

We provided Scale-Sim with detailed information regarding the CNN workload, including the dimensions of each layer, the IFMap dimensions, filter dimensions, and the number of channels for each layer. Scale-Sim leveraged this information to report the clock cycles required to execute each layer, hardware utilization percentage, memory bandwidth, and DRAM index traces for the entire CNN execution. 
We then aggregated the clock cycles required for each layer to determine the total number of clock cycles needed for the entire CNN workload. Our proposed TPU-IMAC architecture enables the execution of each FC layer in just one cycle, and therefore, the overall performance improvement can be calculated by dividing the total number of clock cycles required for the entire workload on the TPU alone by the sum of the clock cycles needed to execute the convolutional layers on the TPU and the clock cycles required to run the FC layers on the integrated IMAC. It is important to highlight that due to the direct connection between the PEs in the systolic arrays and the IMAC, no cycles are wasted transferring data between the systolic array and the IMAC.

Table 2 presents the execution times in cycles while running the CNN on both the TPU and the TPU-IMAC architecture. The results presented in Table 3 reveal a significant improvement in performance when using the proposed TPU-IMAC architecture, particularly for the LeNet model with the MNIST dataset, where we observed a $2.59\times$ performance improvement. Furthermore, improvements ranging between $1.05\times-1.2\times$ were observed while executing other models such as VGG, MobileNetV1, MobileNetV2, and ResNet-18. These variations in performance improvement can be attributed to the size and number of FC layers executed on the IMAC in each model. Specifically, larger sizes and more FC layers in a CNN model tend to result in greater performance improvements when using our TPU-IMAC architecture. These findings demonstrate the potential benefits of our proposed architecture for executing CNN workloads on mobile devices, particularly those with a large number of FC layers.




\section{\textbf{Conclusion}}

In this paper, we presented a novel architecture, namely the TPU-IMAC, which combines the advantages of both heterogeneous mixed-precision and mixed-signal techniques to improve energy efficiency and performance for CNN inference on mobile devices. Specifically, we propose integrating analog IMAC units with digital TPUs to implement FC and convolutional layers of CNN models, respectively. We investigated the architecture-level requirements for efficient realization of the TPU-IMAC architecture and evaluate its potential performance and energy benefits via architecture-level simulations of various CNN models, e.g., LeNet, VGG9, MobileNetV1, MobileNetV2, and ResNet. The simulation results showed that the TPU-IMAC architecture can achieve a performance improvement ranging between $5\%-158\%$ depending on the number and size of FC layers in the CNN workload. Furthermore, our analysis reveals that the proposed TPU-IMAC architecture can significantly reduce memory requirements by $8\%-88\%$, depending on the model. These improvements follow Amdahl's law and are proportional to the ratio of FC layers to convolutional layers. In addition, as IMAC uses ternary weights, we developed an architecture-aware mixed-precision CNN model training methodology to mitigate potential accuracy drops. Results show that this methodology incurs a minimal accuracy drop for CNN models deployed on our TPU-IMAC architecture. In particular, we reported accuracy drops ranging between $0.12\%$ and $0.59\%$ for CIFAR-10, $1.13\%$ for MNIST, and nearly $3\%$ for CIFAR-100 datasets. Our findings highlight the potential benefits of using TPU-IMAC architecture for CNN inference on mobile devices, and provides several opportunities for future work.

\printbibliography


%



\end{document}